\documentclass[final,5p,times,twocolumn]{elsarticle}
\makeatletter
\def\ps@pprintTitle{%
 \let\@oddhead\@empty
 \let\@evenhead\@empty
 \def\@oddfoot{}%
 \let\@evenfoot\@oddfoot}
\makeatother
\PassOptionsToPackage{hyphens}{url} 
\usepackage[hidelinks]{hyperref}
\usepackage{amssymb}
\usepackage{xcolor}
\usepackage{soul}
\usepackage{lipsum}
\usepackage{todonotes}
\usepackage{enumitem}
\usepackage{breakurl} 
\usepackage{longtable}
\usepackage{graphicx}
\usepackage{lscape}
\usepackage{float}
\usepackage{booktabs}
\usepackage[acronym]{glossaries}

\makeglossaries

\newlist{questions}{enumerate}{2}
\setlist[questions,1]{label=RQ\arabic*.,ref=RQ\arabic*}
\setlist[questions,2]{label=(\alph*),ref=\thequestionsi(\alph*)}

\journal{Applied Energy}
\begin{document}
\begin{frontmatter}
\title{Current Practices in Electricy Demand and Charging Scheduling for On-Road Electric Fleet Operations: An Industry-Wide Review}

\author[inst1,inst2]{Joost Commandeur}
\author[inst1]{Bart De Schutter}
\author[inst1]{Neil Yorke-Smith}

\address[inst1]{Delft University of Technology, Delft, The Netherlands}
\address[inst2]{Shell Global Solutions B.V., Amsterdam, The Netherlands}

\begin{abstract}
The electrification of on-road fleet logistics promises improved air quality, lower noise emissions, major climate benefits, increased energy flexibility through the use of locally generated electricity and reduced dependence on imported fuels. However, battery electric vehicles can introduce operational planning challenges not present with internal combustion engine vehicles, including heterogeneous charging speeds, exposure to volatile electricity prices, and scarcity in infrastructure. Managing these complexities requires solutions that balance cost efficiency and robustness, supported by sector coupling between transport and electricity systems. This paper reviews the current state of digital systems for operational decision-making in electric fleet management through a grey literature analysis, drawing on practitioner-oriented sources such as industry reports, company documentation, and technical blogs that reflect real-world practices and developments. We identify key trends and gaps, providing insights to guide future research and development.
\end{abstract}

\begin{keyword}
Electric vehicles \sep Scheduling \sep Forecasting \sep Recharging \sep Microgrid \sep Optimisation \sep Smart Charging \sep Grey Literature Review

\end{keyword}

\end{frontmatter}

\newacronym{bev}{BEV}{Battery Electric Vehicle}
\newacronym{oem}{OEM}{Original Equipment Manufacturer}
\newacronym{ac}{AC}{Alternating Current}
\newacronym{dc}{DC}{Direct Current}
\newacronym{ems}{EMS}{Energy Management System}
\newacronym{cpms}{CPMS}{Charge Point Management System}
\newacronym{fms}{FMS}{Fleet Management Software}
\newacronym{tms}{TMS}{Transport Management Software}
\newacronym{soc}{SoC}{State of Charge}
\newacronym{ocpp}{OCPP}{Open Charge Point Protocol}
\newacronym{bess}{BESS}{Battery Energy Storage System}
\newacronym{bms}{BMS}{Battery Management System}
\newacronym{msp}{MSP}{Mobility Service Provider}
\newacronym{cpo}{CPO}{Charge Point Operator}
\newacronym{vpp}{VPP}{Virtual Power Plant}
\newacronym{openadr}{OpenADR}{Open Automated Demand Response}
\newacronym{ocpi}{OCPI}{Open Charge Point Interface}
\newacronym{glr}{GLR}{Grey Literature Research}
\newacronym{v2g}{V2G}{Vehicle to Grid}
\printglossary[type=\acronymtype, title={List of Abbreviations}]
\section{Introduction}
\label{introduction}
The on-road logistics sector is undergoing a significant transformation to align with global climate objectives, particularly the Paris Agreement \citep{paris2015}, which stresses the importance of substantial reductions in greenhouse gas emissions. To achieve these targets, multiple alternative energy carriers have been explored as substitutes for conventional fuels like diesel and other fossil-based hydrocarbons. These include biofuels derived from renewable feedstocks, low-carbon hydrogen, and electricity \cite{gray2021decarbonising}.\par
Among these options, electrification has gained significant traction in recent years, as reflected in global vehicle sales across various vehicle categories. In 2023, \glspl{bev} accounted for 18\% of new passenger car sales, 26\% of buses, and 47\% of two- and three-wheelers, while battery-electric vans and trucks reached a 4\% market share \citep{bloomberg2024}. The latter figure is expected to increase as all major European truck \glspl{oem}, following their Chinese counterparts, have taken heavy-duty battery-electric models in series production, signalling a commitment to electrification in the heavy-duty segment \cite{volvo2019electric, scania2020electriclaunch, daimler_eactros_2021, renault2023etech, daf2023electricplant, iveco2023pressrelease, man2025electric}. While decarbonization is a primary driver, electrification of logistics is also motivated by a broader set of operational and economic factors. Electricity enables the use of locally generated energy through on-site renewables (e.g. solar power), reducing dependence on imported fuels, increasing energy resilience and enabling cost optimization opportunities through dynamic electricity use.\par
The transition towards electric mobility comes with several advantages, such as the potential to significantly reduce greenhouse gas emissions \cite{verma2022life}. However, the transition also introduces several challenges that must be addressed, which were not present or less prevalent in conventional fleet operations. These challenges can be grouped into three interrelated categories: (i) electricity price volatility, (ii) charging duration and range limitations, and (iii) charging infrastructure and power availability. Digital systems are being designed and implemented to help address these challenges \cite{Plant_Fleetio_EV_Fleet_Management_2024}.\par
This paper provides a structured review of current industry practices in digital systems and processes that support operational decision-making in electric fleet mobility management, addressing the unique challenges introduced by electrification. It is intended for two primary audiences: academic researchers seeking real-world insights to inform optimization and control strategies research, and industry practitioners/solution providers aiming to understand current capabilities, integration gaps and standardisation opportunities. By mapping current practices and emerging directions, the paper offers a comprehensive view of the digital ecosystem and its key players, enabling the identification of opportunities for digital innovation and interoperability standards. Through this synthesis, we aim to bridge theory and practice, offering actionable recommendations for future research to accelerate the research and development of robust, integrated solutions for site energy management, charging coordination, and fleet scheduling.\par
The following research questions form the basis:
\begin{questions}
    \item Which operational decisions related to electric fleet management are currently automated and which decisions remain manual or unsupported?
    \item Which industry actors provide solutions for which kind of automated decision making, what functionalities do they offer and what do these systems optimise for?
    \item To what extent do existing solutions enable automation and integration—ranging from decision support to fully optimised operational plans?
    \item What types of data and information sources (e.g., vehicle telematics, energy prices, grid constraints) are incorporated into these solutions for decision-making?
\end{questions}\par
This paper is organised as follows. Section 2 elaborates on the scope of this review; it outlines the challenges introduced by fleet electrification and describes the relevant digital systems and their associated actors, concluding with a categorization framework for trend analysis. Section 3 outlines the systematic approach used to identify and assess market trends. Section 4 presents the findings of the industry-wide review, including observed trends and patterns. Section 5 discusses the limitations of this study. Finally, Section 6 summarizes key insights and provides recommendations for future research.\par

\section{Operational Context and Scope of Review}
This section delineates the scope of the review, namely operational decision-making systems for electric fleet logistics. It outlines key electrification challenges, introduces relevant industry actors and digital tools, and presents a classification framework to structure the analysis.
\subsection{Scope}
The scope of this review is specific to professional logistic fleets which are centrally managed under coordinated schedules. It excludes ad hoc or unscheduled vehicle use, such as field service fleets without centralised planning. The typical scope is therefore on first, middle and last mile transport. The focus is on digital systems and aspects of digital systems that facilitate operational decision-making impacting and deciding the provisioning of electricity and the charging process as these are the novel challenges introduced by electric vehicles. Vehicle routing and planning is within scope, but only as related to decisions impacting the timing of the charging session and the amount of electricity required as these are specific to electric vehicles. This review excludes automation pertaining to strategic and tactical decisions, such as optimal sizing and location models for investments, as well as automation that does not directly impact operational decision-making, such as logistics demand forecasting. These tools can aid decision-making processes but do not suggest decisions and therefore the technical details of these systems are not within the scope of this review.
\subsection{Challenges}
Electrification of fleet logistics introduces distinct operational challenges absent in conventional fuel systems. This subsection outlines these challenges and highlights how digital systems are increasingly leveraged to manage the resulting complexity.

\subsubsection{Electricity price volatility and source availability}
The supply chain for electricity differs fundamentally from that of conventional fuels like diesel. It is shorter, with production and consumption occurring within the same region, and significantly more volatile in price over a short time horizon \citep{pavlik2025renewable, CBS_80416ENG_misc, EC_DGENER_WeeklyOilBulletin_misc}. Electricity can be produced locally through solar panels and wind turbines, or centrally via offshore wind farms, gas turbines, and hydropower stations. The flexibility to produce electricity more locally has enabled the rise of microgrids, where generation and consumption occur at the same location. While microgrids offer cost and resilience benefits, they also introduce operational complexity \citep{shahzad2023possibilities}.\par
Electricity grids, both local and national, need to be in constant balance between supply and demand since grid lines lack inherent storage capacity. Electricity can be transported over the grid lines but it needs to be converted in other forms of energy before it can be stored, such as electrochemical (e.g., \gls{bess}), mechanical (e.g., pumped hydro), thermal (e.g., molten salts), or chemical (e.g., hydrogen) energy \citep{gur2018review}. In an \gls{ac} grid, which is used at the national level in most countries, any imbalance between supply and demand results in fluctuations in the frequency of the electricity. Both supply and demand equipment are designed to operate at a specific frequency, and deviations from this frequency can cause damage to the devices \citep{kirby2003frequency}. The supply side of the electricity market is increasingly volatile as more solar and wind power generation is added \citep{clo2015merit}. These renewable sources are inherently intermittent, as their output depends on fluctuating environmental conditions such as solar irradiance and wind speed.\par
Maintaining grid stability, matching demand and supply, can be achieved through three main approaches: adjusting supply output to match demand, leveraging electricity storage to bridge supply-demand gaps, or modifying demand patterns through demand-side response to follow supply \citep{salman2022review}. On a local level, microgrids rely on \glspl{ems} to forecast and monitor demand and supply and effectively direct electricity streams, potentially also to local storage. The \gls{ems} coordinates between the three levers (controlling supply, demand and storage) to maintain local grid stability as a single decision maker \citep{shahzad2023possibilities}.  At regional or national scales, where there are multiple suppliers and consumers who are the decision makers, coordination mainly occurs via electricity markets. These markets are structured to enable market participants to cost-effectively coordinate among these three coordination options. The main coordination instrument is the market price which encourages or discourages production, demand and storage. These electricity prices are increasingly volatile due to the integration of more intermittent electricity supply sources, e.g. the price per hour in 2024 in the Netherlands, a country with high penetration of renewable electricity production \citep{CBS_2025_RenewablesHalf}, fluctuated between ±10 and ±120 euro/MWh \citep{tennet2024marketreport}. To put this in perspective, for a truck driving a typical 100.000 km per year at ± 1 kWh/km this would mean either a 1.000 euros or 12.000 euros on electricity cost.\par
As a result, the logistics sector, when transitioning to electric vehicles, may face increased exposure to fluctuating energy prices on an hourly time scale as compared to conventional fuels which can result in additional operational planning complexity. Traditionally, on-road logistics companies did not need to account for energy price fluctuations in their operations and the planning thereof, as prices for fossil fuels remained stable over the typical maximum planning horizon of 2-5 days \citep{CBS_80416ENG_misc}. Exposure to fluctuating electricity prices depends on the type of electricity contract that the logistics companies have. Electricity retailers can offer stable prices at a premium. The risk and potential benefit of the fluctuating electricity prices then sits with the electricity provider.\par
When companies in the logistics sector opt for a microgrid solution to either manage limited grid connection or guard against grid price volatility, they will need \gls{ems} solutions to manage the power flows effectively. The electricity demand resulting from their transport operations will then need to be integrated into these \gls{ems} solutions to allow the system to efficiently coordinate power flows, adding integration requirements on electricity demand schedules and potential additional operational complexity.

\subsubsection{Charging duration and range limitations}
A second challenge stems from the fact that \glspl{bev} have a more limited range as compared to their ICE counterparts \citep{IEA_GlobalEVOutlook_2025} and charging a \gls{bev} takes longer charging then the refuelling time of an ICE vehicle. Traditional fuel dispensers, which supply diesel or gasoline, come in two forms: standard or high speed. The standard fuel dispensers supply fuel at a rate of 20 to 40 litres per minute, while high-speed fuel dispensers supply 60 to 80 litres per minute \citep{Elaflex_ZVA25_HiFlo_Datasheet_2008}. Considering a typical consumption rate of 27 liters per 100 kilometres \cite{ICCT2023HDVDecarb} for 40-ton trucks, a high-speed fuel dispenser provides 220 to 300 kilometres of range per minute. In the case of charging a \gls{bev}, the speed of charging (i.e., charging power) is determined by both the maximum charger speed as well as the maximum charging speed of the battery in the vehicle. There are various types of chargers on the market ranging from slower 3 to 44 kW \gls{ac} type chargers, to 50 kW overnight \gls{dc} chargers to 400 kW high-powered top-up \gls{dc} chargers. The latest development is the new MCS standard which facilitates charging speed of 1000+ kW \cite{CharIN_MCS}. Taking a typical consumption of 1 kWh per kilometre \cite{ICCT2023HDVDecarb} for the same heavy duty truck this means \gls{ac} type chargers supply electricity at a rate of 0.05 to 0.73 kilometre per minute, overnight \gls{dc} chargers at 0.83 kilometre per minute, and high-powered top-up \gls{dc} chargers at 6.67 to 16.67+ kilometres per minute. As can be seen, this is orders of magnitude lower than the speed at which conventional fuels can dispense "kilometres". Having a high-speed charger is also not a guarantee of faster charging speeds. The charging speed is determined by the battery in the vehicle. The battery has a maximum charging capacity and during a charging session this maximum charging speed can be reduced by the \gls{bms} to improve battery health by preventing overheating. Throttling down the charging speed by the \gls{bms} results in a so-called charging curve, which indicates what the maximum charging speed at a certain \gls{soc} of the battery is. The charging curve is influenced by multiple factors and therefore spans a range of possible profiles \cite{adaikkappan2022modeling}. This longer and more uncertain charging behavior of electric vehicles needs to be accounted for in operational planning.

\subsubsection{Infrastructure and power availability}
The last type of new challenge with \glspl{bev} relates to the availability of infrastructure and power for charging. The on-road logistics sector has not come to a conclusion yet on the ideal charging hardware configuration both in terms of the location of these chargers to support the fleets as well as the combination of number of chargers per vehicle and type of chargers. Similar to the conventional fuels, the location of charging can happen both en-route at public charging facilities as well as at private locations from which the fleet operates, so called \textit{depots}. Because en‑route charging increases journey time and incurs driver costs, unlike depot charging when vehicles are parked, and because depots can integrate low‑cost local renewable generation, it is expected that most electricity demand will be met at depot locations. \citep{RefaEVS38_2025}. One of the potential downsides of depot charging is the lower utilisation rate of the charging infrastructure as it is dedicated to a single fleet, resulting in higher costs. A one-to-one pairing of chargers to vehicles on a private site for a fleet results in no operational constraints when planning the charging sessions as chargers will always be available, but a high number of chargers results in high costs. Therefore it is likely that a trade-off will be chosen between the number of chargers and scheduling the use of them efficiently. This needs to then be integrated into the overall fleet planning, resulting in additional complexity. For the public charging sites a similar infrastructure scarcity problem arises. Given that charging sessions take longer than refuelling a vehicle, the situation can arise that the charger is already occupied and that it takes a significant amount of time for the charger to become available again.\par
Next to a potential scarcity in available (high-powered) chargers there is a potential scarcity in available power on the site. To accommodate the required power for the electric fleets many depots sites as well as public charging sites will have to increase their grid connection capacity as they are not designed for the increasing power demand caused by electric fleets. Depending on the country and region these grid upgrades can take a significant amount of time (even up to several years) \citep{Hacker_Truck_Depot_Charging_2025}, resulting in temporal scarcity of power. In the long term it is also the question if sites will have a grid connection that is designed to accommodate peak power or that power buffering in the form of \gls{bess} will be more cost-effective in combination with a more moderate grid connection size. When there is power scarcity, especially in the case when there is power buffering, these limitations need to be accounted for in the operational planning of the charging sessions as ignoring them would result in longer than anticipated or failed charging sessions. 

\subsection{Electric fleet actors and their digital systems}
The digital systems, such as charging and fleet management software systems, that are the topic of this paper support humans/decision makers in their decision making by suggesting and/or enforcing solutions. The following outlines which humans/entities are involved in operational planning for \gls{bev} fleets and which digital systems these actors use. We will limit the scope to actors that make or support the operational decision making in electric fleet operations. Some actors can be the same person/entity. For context, we also provide a brief non-exhaustive list of the digital systems that supply input data rather than decision suggestions at the end of each actors section.
\subsubsection{Site manager}
The site manager is responsible for the safety and cost effectiveness of site operations. Electric/energy operations on a site concerns managing all power flows. This entails managing the grid connection and staying within bounds of the grid capacity contract. It also concerns managing the (scheduled) operations of assets on the site that are not directly related to the fleet, such as warehouse climate control, production assets, convenience stores (public charging), and potentially on-site energy storage through e.g. batteries or heat storage.\par
The digital system that this actor uses is the \gls{ems}. The \gls{ems} typically links asset management software, building management software and \gls{bms} for \gls{bess} under one unified control scheme from a power perspective. The \gls{ems} can include controlling electricity flows, ramping up and down power levels of heating and cooling equipment, managing the inflow of solar PV produced electricity, and managing charging and discharging of stationary battery storage.\par
Some of the digital support systems are digital twins of assets, forecasting software for solar irradiation, temperatures, and electricity market prices. 
\subsubsection{Fleet dispatcher}
The fleet dispatcher is tasked with creating a dispatching schedule for the fleet of vehicles to serve the logistics demand. This entails assigning which vehicle will drive which route and when, while considering operational constraints such as driver availability, customer service windows, service contract obligations and vehicle limitations like range and load capacity.\par
The main digital systems used are \gls{fms} and \gls{tms}. These terms are not strictly defined and often overlap. Typically, \gls{fms} focuses on vehicle-level planning, starting from predefined full truck loads, whereas \gls{tms} begins at the shipment level—pallets, packages, or volumes—and aggregates these into full truck loads before assigning them to vehicles. Depending on the customer’s operational setup, both systems may coexist. The optimisation engine that determines the best assignment of pallets, packages, volumes, or trucks to routes can be either integrated into these platforms or added as a plug-in.\par
Telematics systems blur the line between decision making and supporting software. It is the technology that combines telecommunications and informatics to transmit, receive, and store vehicle-related data such as location, speed, fuel or energy consumption, and battery state of charge. Some telematics providers extend their functionality beyond monitoring to include route planning and optimisation, leveraging historical operational data to match routes with specific vehicles or drivers based on performance characteristics.\par
Supporting software systems include workforce scheduling tools, enterprise resource planning platforms, customer demand portals, traffic forecasting and electricity consumption estimators, which complement scheduling decisions but do not directly optimize routes. 
\subsubsection{Driver}
The driver operates the vehicle and executes the plan that has been made by the dispatcher. Depending on how a company has structured its decision making process it could be that the driver also controls part of the operational decision making. These driver decisions could be either for pre-defined decisions, such as break times or in the event that changes need to be made in the original plan, then sometimes the driver has to find the best possible alternative.\par
The drivers make use of so-called driver apps, which are typically part of the \gls{tms} or \gls{fms} solutions. These applications give real-time information on the route that they are driving and also provide a communication means between the fleet dispatcher and the driver.\par
Supporting software for the driver apps are vehicle telematics to understand state of charge and consumptions rate, traffic data management systems, and book and charge systems to reserve a charging spot at a public charging facility. 
\subsubsection{Charge point operator}
The \gls{cpo} is the entity responsible for operating the charge points and ensuring their availability and functionality. This role can overlap with the site manager but does not necessarily have to. The primary digital system used by the \gls{cpo} is the \gls{cpms}, which governs charger operation and provides the interface for monitoring and control.\par
The complexity of \gls{cpms} requirements depends on the type of site and its operational context. Public charging sites are inherently multi-party environments, serving vehicles from different operators, even passenger vehicles, and therefore requiring robust functionality for authentication, billing, and sometimes reservation management. Private depot sites, by contrast, can be either single-party or multi-party. Single-party depots typically require only basic \gls{cpms} capabilities, as there is no need for cross-fleet billing or authentication. In these cases, the system may simply ensure safe operation and deliver maximum charging power to vehicles. \par
However, when operators aim to implement advanced services—such as scheduled charging, smart charging, or dynamic load management—the \gls{cpms} must become more sophisticated, even for single-party sites. These features require coordination with site-level energy constraints, vehicle priorities, and external factors like electricity tariffs, moving beyond the “charge as fast as possible” approach.\par
Multi-party private sites can be chosen to reduce infrastructure costs and improve asset utilisation. In such cases, \gls{cpms} must support booking systems, authentication, and cost attribution across different operators, while respecting site power limits and contractual obligations. This introduces interoperability challenges with \gls{ems} and fleet scheduling platforms, as charging decisions can no longer be made in isolation. 
The supporting software for the \gls{cpo} is book and charge software such that the operator can block chargers to reserve them for incoming users. 
\subsubsection{Mobility service provider}
\glspl{msp} facilitate charging access for their customers to roaming networks of charge points and facilitating payments to the charge point operators.\par
The digital systems that the \glspl{msp} use is price settlement software. By determining a price point they can incentivize their customers to go to specific charging point, potentially also at specific times.\par
The supporting software that \glspl{msp} use is the price communication software of \gls{cpo}s.
\subsubsection{Energy provider/Aggregator}
The energy provider, or electricity contract company, provides the electricity to end customers, which in this case are logistics companies. These providers typically aggregate a large number of customers into a single trading position in the electricity markets. These companies estimate demand and buy electricity for their customers. They can also trade the potential demand flexility of their customers on ancillary markets. For this they will need to make agreements with their customers on delivery service level agreements.

\subsection{Categorization of digital operational decision support systems}\label{categorization-levels}
There are several levels on which decisions are made regarding charging and electricity management that all influence the availability, price, and demand for fleet electricity. To be able to make a structured review of industrial practices a categorization is defined that specifies these different operational levels. Software systems could span multiple of these levels at the same time. We identify three levels:\par
\begin{figure}[h]
    \centering
    \includegraphics[width=0.5\textwidth]{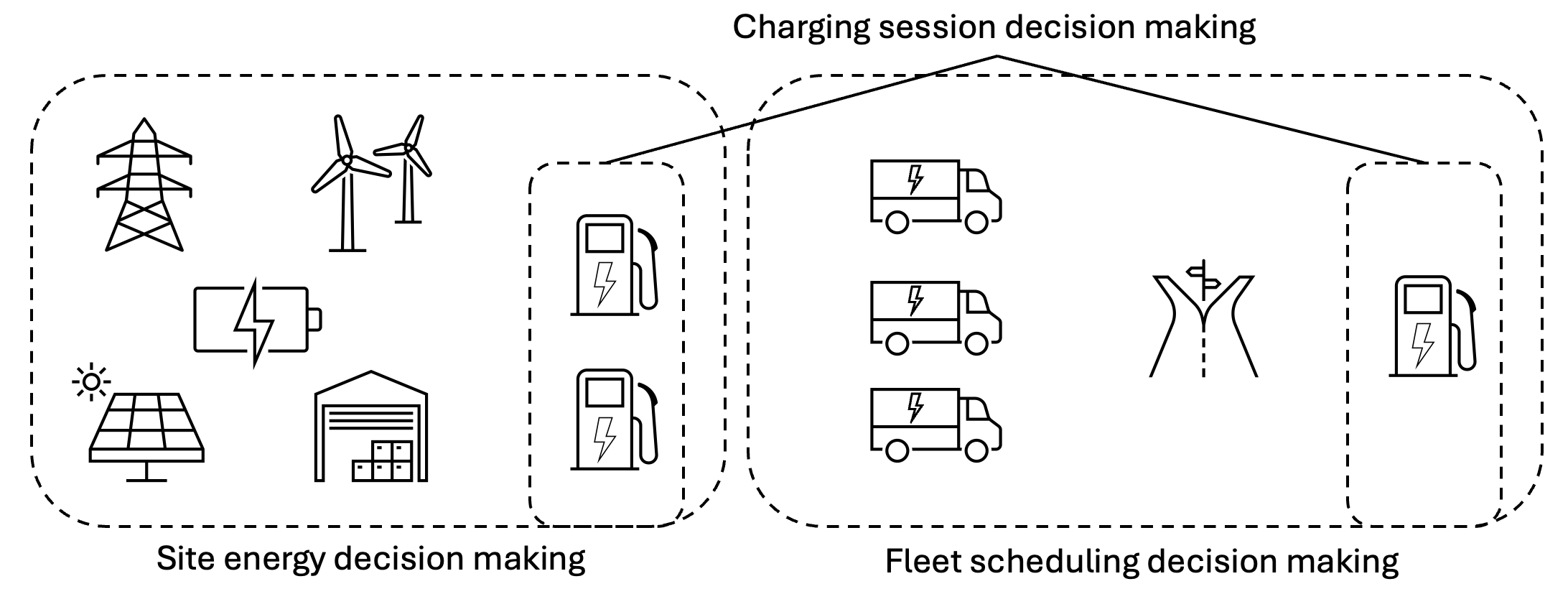}
    \caption{Operational decision levels in electric fleet management}
    \label{fig:levels-graph}
\end{figure}

\begin{enumerate}
    \item \textbf{Site energy decision making}\\
    Entails the decision making processes related to energy resources at the physical location from where the fleet operates. Key activities include: energy storage management, grid connection management, and site load management. 
    \item \textbf{Fleet scheduling decision making}\\
    Involves the planning and optimisation of fleet operations to ensure efficient and timely service. Key activities include route (re-)optimisation and dispatch planning. 
    \item \textbf{Charging session decision making}\\
    Focuses on the efficient management of vehicle charging sessions to ensure availability and to minimize downtime. Key activities include smart charging, load management, and pricing strategies.
\end{enumerate}
This categorization will be used in the remainder of this industry-wide review to structure the approach and findings. Figure \ref{fig:levels-graph} presents a graphical depiction of the levels for clarity. Table \ref{table:actors-table} indicates which actor is active in which level of operational decision making.
 
\begin{table}[h!]
\centering
\begin{tabular}{lccc}
\toprule
 & Site level & Fleet level & Charging level \\
\midrule
Site manager    & x &   & x \\
Fleet dispatcher &   & x & x \\
Driver          &   & x & x \\
\gls{cpo}             &   &   & x \\
\gls{msp}             &   &   & x \\
\bottomrule
\end{tabular}
\caption{Actors in operational decision levels}
\label{table:actors-table}
\end{table}

\section{Methodology}
This section elaborates on the concept of a grey literature research and describes the methodologies used to find the answers to the research questions. To assess the current state of practice in digital systems for decision making in the on-road electric fleet sector, several methods were employed.
\subsection{Grey literature review}
A grey literature review examines non–peer-reviewed sources from practitioners, e.g. industry reports, vendor documentation, technical blogs, white papers, and standards. This approach as described by \citet{garousi2019guidelines} captures current practitioner insights and market realities that traditional reviews may miss. We adopt a \gls{glr} because the digital systems in the electrified logistic space evolve rapidly, and academic publications may not always capture the most recent industry developments. Practitioner-oriented sources therefore provide timely insights into current capabilities, integration challenges, and emerging directions.
\subsection{Data collection}
An internet search was conducted to review the websites of key market participants, identifying the features and capabilities they advertise. This provided insights into current offerings and strategic focuses. Interviews were conducted with a selected number of companies within the sector to understand the directions in which they are developing their software systems and the challenges they face. The keywords to use in both search activities were distilled from expert interviews from within the markets, consultation of Large Language Models and through snowballing effects during the search process itself.\par
\citet{kliman2015location} showed that search results are dependent on the IP address of the machine requesting the information from the search engine. This results in a so-called \textit{filter bubble}. The topic of this article, and therefore the search queries, is strongly linked to the \textit{local search queries} and therefore prone to strong location dependence \cite{kliman2015location}. The search engine used in this research allowed to change the location of the search query request, which was set to Amsterdam, Bangalore, Hong Kong, and Houston to cover the areas of interest for this article. All other personalization settings were shut off to the extent that the search engines allowed. 
For charging session decision making the following keywords were used: \textit{Charge Point Management Software}, \textit{Electric Vehicle Charging Software}, and \textit{Smart Charging Solutions}.\par
For site level electricity decision making the following keywords were used: \textit{Business Electric Fleet Energy Management Software} and \textit{Battery and Solar Energy Management for Fleet}\par
For electric fleet scheduling decision making the following keywords were used: \textit{Fleet Management System for Electric Vehicles}, \textit{Electric Vehicle Route optimisation} and \textit{Electric Vehicle Fleet Dispatch Software}\par
As online search engines can provide millions of hits against a search query it is impractical to do an exhaustive search through all results. As described by \citet{garousi2019guidelines} on conducting grey literature reviews, it is important to define the stopping criterion. In this research the stopping criterion has been defined as reaching the point where no new information is found. Going through the search engine results the quality of the hits diminishes, when the point was reached that the hits did not add any new information the search was stopped. As this study has qualitative questions and not quantitative, the number of results found  is less important than the quality of results and the ability for the results to give insights in the developments in the field.
\subsection{Selection criteria}
To ensure the relevance and quality of the data collected from the search activities, specific selection criteria were established:
\begin{itemize}
    \item The primary criterion was that the digital systems had to be directly connected to operational electric fleets. Especially for site level management there are many digital systems called energy management systems that do not include any specific fleet related features and only focus on solar photovoltaics and battery storage systems. There is also a multitude of solutions for home energy management systems that do include electric vehicle charging, but since these are focused on consumer passenger vehicles these are also excluded as this article focuses on operational fleets and not on individual consumer passenger vehicles. Lastly all digital systems that are not electric vehicles specific are also excluded.
    \item In addition to relevance to the subject, the results found also need to be recent as this article focuses on the current practices and development directions. Given that he market is evolving fast, only results that are no older than two years were considered (January 2023 and later).
    \item Only results that were available in English were considered for this research.
    \item Credibility of the source is the last selection criterion used. Only market parties with actual customers or customers were considered. This ensures the practical applicability of the solutions advertised.
\end{itemize}
By utilising the search terms and applying the selection criteria, a total of 91 industry solutions have been found: 16 solutions were found for site energy decision making \cite{nuvve_platform, eaton_brightlayer_energy, Spirii2026, HitachiZeroCarbon2026, RiDERgy2026, Synop2026, PowerON2026, FutureEnergy2026, BluwaveAI2026, ABB2026, Ampcontrol2026, Ampowr2026, Honeywell2026, iWell2026, Schneider2026, Ampfox2026}, 48 solutions for fleet scheduling decision making \cite{MichelinConnectedFleet2026, LocoNav2026, TrakM82026, KiaPBVFMS2026, Trimble2026, GPSInsight2026, XotrucksXosphere2026, FYND2026, SmartTrak2026, GridLine2026, Volteras2026, ZeroMission2026, RoadCast2026, Fryte2026, eMotionFleet2026, TeletracNavman2026, BetterFleet2026, OptiBus2026, MoevAI2026, Tenix2026, Katsana2026, Chetu2026, ZonarSystems2026, MagentaMobility2026, MakeMyDay2026, AutoWiz2026, FleetX2026, TrinetraWireless2026, Trackobit2026, Loginext2026, GoMotive2026, FarEye2026, NextBillionAI2026, Iveco2026, PTVLogistics2026, Ortec2026, Navixy2026, StandardFleet2026, MatrackInc2026, FleetBoard2026, ConsatTelematics2026, Webfleet2026, BlueArrowTelematics2026, TargaTelematics2026, Geotab2026, Scania2026, Ituran2026, Samsara2026}, 24 solutions for charging sessions decision making, both \gls{cpms} and \gls{msp}  \cite{Driivz2026, Ampeco2026, Monta2026, OceanEV2026, Greenflux2026, LastMileSolution2026, Vitra2026, Current2026, SwitchEV2026, Emabler2026, ChargePoint2026, EOCharging2026, ChargeLab2026, EVConnect2026, TridensTechnology2026, ChargePanel2026, Reev2026, Evoltsoft2026, ClenergyEV2026, Vialumina2026, Deftpower2026, Plugsurfing2026, Bluedot2026, dhemax_website}, 3 general software development companies supplying solutions to all three categories \cite{Codibly2026, Tekmindz2026, ZealousSystem2026} and 1 mapping service which integrates EV constraints in scheduling routes \cite{Here2026}. 

\subsection{Analysis approach}
For each decision level, as specified in Section \ref{categorization-levels}, a table is constructed based upon all features mentioned by the market parties. For conciseness these tables are level-specific. Each identified solution will be systematically presented by indicating the presence or absence of the feature indicated by the column header. If a feature has been announced but not yet available or implemented, this is also specifically indicated in the table. The creation of the table itself is part of the result of this article as it gives an overview of the advancements in this field. The full table is part of the appendix to this paper, the general insights will be presented in Section \ref{overview-of-practices}\par
The categorical insights are then used for a comparative analysis between solution within each level as well as differences between the levels themselves which will be shown in Section \ref{commonalities-and-differences}. Lastly the trends and patterns which have been identified will be shown in Section \ref{trends-and-patterns}.
\section{Findings}
This section synthesizes insights from the industry-wide review of digital systems supporting operational decision making for electric fleet logistics. The findings are organised across three operational layers—site-level energy management, fleet scheduling, and charging session management—followed by a cross-level comparison and emerging trends. A detailed table overview of the findings can be found in the tables in \ref{sec:findings_tables}.
\subsection{Overview of practices}\label{overview-of-practices}
\subsubsection{Site-level energy decision making}
\glspl{ems} are well-established for traditional site-level tasks, including solar PV forecasting and dispatch, \gls{bess} scheduling, grid connection limit enforcement, and tariff-based optimisation. These capabilities enable sites to manage energy flows effectively. However, fleet-specific integration remains rare. Most \gls{ems} platforms do not ingest operational fleet data, such as \gls{soc} or charging priorities, which limits their ability to optimise charging alongside other site loads. \glspl{ems} typically communicate using the modbus protocol, whereas the standard communication protocols for chargers for more rich charging data are \gls{ocpp} and \gls{ocpi}.\par
Some notable exceptions are ABB DynovaPro \cite{ABB2026} and Spirii \cite{Spirii2026}, which combine \gls{ems} and \gls{cpms} functionalities, enabling ingestion of \gls{ocpp} data for charger-vehicle mapping and improved coordination. Two other notable exceptions are that of Synop.AI \cite{Synop2026} and Ampcontrol \cite{Ampcontrol2026} which both integrate with telematics for real-time updates on arrival time and required electricity of the vehicles. These integration allow the systems to align charging sessions with site-level constraints, but such solutions are outliers in the current market.\par
Where \gls{ems} and \gls{cpms} coexist on a site, integration typically occurs through API/OCPP interfaces, but bi-directional control and communication is not yet widely supported. \glspl{ems} hold information on \gls{bess} flexibility, solar generation, and non-fleet loads, while \glspl{cpms} manage charging bookings and priorities. Without a unified optimisation engine, decisions are made sequentially and iteratively, resulting in optimisation problems with incomplete information across systems (partially observable). This can lead to suboptimal outcomes, as each system optimises based on incomplete information, or requires iterative adjustments between separate optimisation engines.\par
Multi-fleet integration at the \gls{ems} level was not observed in practice. \gls{cpms} platforms, however, commonly support multi-fleet operations on public charging networks, offering features such as slot booking and driver authentication. \par
Key observations include:
\begin{itemize}
    \item \gls{ems} platforms are mature for traditional energy tasks but lack fleet-specific features.
    \item Integration between \gls{ems} and \gls{cpms} platforms is emerging but limited to a master-slave set-up, without iterations to find global optima.
    \item Co-optimisation across \gls{ems} and \gls{cpms} remains aspirational, hindered by fragmented data and incompatible objectives.
\end{itemize}

\subsubsection{Fleet scheduling decision making}
Fleet scheduling solutions are the least mature when it comes to integration toward the other two layers. Interviews revealed that most logistics operators own only a handful of \glspl{bev}, often deployed for marketing purposes, emission zone compliance, or limited pilot testing rather than full operational integration. Consequently, \glspl{bev} are scheduled conservatively, with range restrictions and minimal optimisation for charging constraints. In some cases, fleets deliberately assign \glspl{bev} to fixed routes within emission-restricted zones, which further limits utilisation flexibility.\par
Current planning tools focus mainly on monitoring \gls{bev}, such as \gls{soc} and consumption tracking, and rarely integrate charging session planning. The only observed EV-specific features are range restriction and charge point visibility on the planning map for human-in-the-loop planning, while advanced capabilities, such as charge slot reservation integration, and electricity price-aware planning, are still mostly on vendor roadmaps with notable exceptions being . More advanced, typically machine-learning based, vehicle consumption models are beginning to appear, accounting for variable energy use across routes and conditions, but these are not widespread. Some examples of this are the consumptions model from \citet{Ortec2026} and \citet{chargetrip} or \citet{inceptev2026} which was acquired by \gls{fms} company \citet{GoMotive2026}. These consumption models are a stepping stone to more automated decision making.\par
Both BEV‑native tools and legacy FMS expanding into BEV support are emerging. With BEV adoption rising gradually, systems must either handle mixed diesel–BEV operations within one platform or be run in parallel, which increases operational complexity. A way for traditional \gls{fms} vendors to extend their \gls{bev} capabilities is by integrating planning modules from specialised providers such as \citet{PTVLogistics2026}, \citet{FarEye2026}, and \citet{NextBillionAI2026}. It is noteworthy that observed developments in EV-specific functionalities are concentrated within \gls{fms} rather than \gls{tms}. While \gls{tms} platforms primarily focus on shipment-level aggregation and routing, they rarely incorporate charging logic or \gls{bev} constraints. From the top 10 \gls{tms} solutions, as presented by Gartner \citep{Gartner_TMS_PeerInsights_2025}, only one \gls{tms} provider showed development on EV specific features. In contrast, \gls{fms} vendors are beginning to introduce features such as range-aware planning, albeit in early stages. This divergence suggests that electrification pressures are currently driving innovation closer to vehicle-level planning rather than shipment-level orchestration.\par
Telematics integration is emerging, primarily in \gls{cpms} rather than scheduling tools. For example, Webfleet \cite{Webfleet2026} enables \gls{cpms} to anticipate arrival state-of-charge and timing during execution. Telematics systems are increasingly leveraged not only for real-time monitoring but also for analysing historical operational trends. These insights enable data-driven recommendations on optimal charging locations and times, supporting planners in aligning route assignments with charging opportunities and minimizing downtime. Such analytics are most effective in operations with high predictability, where recurring patterns allow structural adjustments to routing and charging strategies. In contrast, for fleets with highly variable or diverse operations over time, the benefits of telematics-driven charging suggestions diminish, as irregular patterns reduce the reliability of trend-based recommendations. Telematics also enables remote control of charging sessions (e.g., starting or stopping charging), giving fleet operators greater control over vehicles charging at third‑party or non‑operated sites. This however can cause conflicts with the smart-charging goals that the local \gls{cpms} might have.\par
Planning workflows are predominantly planner-in-the-loop, with no evidence of closed-loop automation. In fact, many fleets still rely on Excel-like tools and experience-based planning, underscoring the sector’s slow adoption of digital solutions. Vendors acknowledge that the complexity introduced by \glspl{bev} may act as a catalyst for automation adoption, but this transition is still in its infancy.\par
A special case within fleet operations is that of bus operations. The bus sector is significantly further ahead in electrification, partly due to stronger government involvement, and consequently its decision-support systems are more mature. Solutions such as \citet{OptiBus2026} and \citet{Tenix2026} illustrate what future capabilities for other fleet types may look like, though bus operations differ fundamentally and require distinct optimisation approaches. Unlike freight or service fleets, bus operations follow highly predictable schedules with fixed routes and timetables. This reduces the need for continuous route optimisation. Decision-support systems for buses therefore prioritize depot charging coordination and energy cost optimisation rather than dynamic dispatch planning. Standards such as VDV463 \cite{VDV463_ShortIntro_2025} are commonly used in this domain for the communication between fleet operations and depot operations, but their focus on timetable adherence limits applicability to more flexible fleet types. An example of integration between fleet and depot operations is that of \citet{OptiBus2026} and \citet{dhemax_website}.\par
New integrator platforms, such as \citet{Fryte2026} together with \citet{hubject_website}, are emerging to link fleet operators, charging infrastructure, and data streams, offering automated capabilities like route planning and book‑and‑charge. As these ecosystems mature, they lessen the need for direct operator–CPO integrations. These bookings are using the \gls{ocpi} booking module.\par
Key observations include:
\begin{itemize}
    \item Fleet scheduling tools lack integrated charging logic.
    \item EV-specific features are minimal; most advanced capabilities remain conceptual.
    \item Market inertia and low \gls{bev} penetration are major barriers to innovation.
\end{itemize}

\subsubsection{Charging session decision making}
\gls{cpms} platforms represent the most mature layer of operational decision making. Common capabilities include dynamic load balancing, priority queues, price-based optimisation, and reservation systems. Advanced features such as charging demand forecasting are emerging, enabling \gls{cpms} to predict energy needs without direct fleet schedule integration—though this approach assumes highly predictable behavior and risks downtime under unforeseen conditions. \par
Larger \gls{cpms} providers initially focused on supporting public \glspl{cpo} at highways and other public locations. They are now expanding into private depot environments, which shifts their approach from predicting charging behavior to ingesting planned behavior. The move towards the private space also results in smart-charging capabilities being added as the dwell time of the vehicle is typically longer at private locations. \par
Fleet schedule ingestion by \gls{cpms} is still nascent, typically via manual uploads (CSV/Excel) or basic UI-based priority settings. \gls{cpms} systems respect site power limits and, in some cases, integrate with solar forecasts and power metering for dynamic load management. Integration with \gls{bess} remains limited. Decision authority is largely reactive, executing operator-defined policies rather than proposing alternative operational modes. Vendors such as Driivz \cite{Driivz2026} and Monta \cite{Monta2026} stand out for advanced functionality, such as \gls{openadr} integration for demand side response in electricity markets, and early moves toward predictive analytics.\par
Real-time fleet information ingestion through telematics is an advanced feature that we see being implemented with the more advanced \gls{cpms} providers such as \citet{ChargePoint2026}. \par
Many \gls{cpms} providers are developing their offering for \gls{v2g}, which allows vehicles to return electricity back to the grid, with a notable example being LastMileSolutions \cite{LastMileSolution2026}. The pilots in this space are mainly focused on non-scheduled fleets of vehicles, typically passenger vehicles on parking lots. For more professional fleet which work according to a set schedule some more integration with the schedule of the vehicle would be required.   
Key observations include:
\begin{itemize}
    \item \gls{cpms} platforms lead in smart charging innovation but lack deep integration with \gls{ems} and fleet scheduling.
    \item Predictive features are emerging but not widely deployed.
    \item Decision making remains reactive; proactive optimisation is a future aspiration.
\end{itemize}

\subsection{Commonalities and differences}\label{commonalities-and-differences}
Across \gls{ems}, \gls{cpms}, and fleet scheduling, several common building blocks exist: tariff feeds, available power data, and fleet telematics (\gls{soc}) are critical for optimisation at all levels. \gls{cpms} requires additional administrative data for billing and invoicing.
Maturity varies significantly:
\begin{itemize}
    \item \gls{cpms} is the most advanced, driven by its proximity to real-time charging operations and lack of legacy constraints.
    \item \gls{ems} is established for traditional energy tasks but slow to adopt fleet-specific features.
    \item Fleet scheduling lags behind, constrained by low \gls{bev} adoption and conservative operational practices.
\end{itemize}

Interoperability remains limited. While \gls{ocpp} is widely used for charger communication, no standardized protocol exists for \gls{ems}-\gls{cpms}-\gls{fms} integration. VDV463, used in bus fleets, is insufficient for optimisation and alternative proposal workflows, a gap acknowledged by its authors \citep{VDV463_ShortIntro_2025}. Current integrations are largely bespoke, creating barriers to scalability although connecting platforms are emerging. The integrated planning across both public and private charging sites also remains a dormant area which requires stronger collaboration between all parties. 

\subsection{Trends and patterns}\label{trends-and-patterns}
Three major trends emerge:
\begin{enumerate}
    \item Convergence and Partnerships: \gls{cpms} vendors are actively expanding into more escheduling and \gls{ems} domains, often through partnerships rather than in-house development. \gls{ems} providers are adding \gls{cpms} capabilities, while the fleet scheduling domain is starting to built more partnerships both towards \gls{cpms} and companies with dedicated EV planning modules.
    \item AI and Predictive Analytics: Machine learning is being applied to \gls{soc} estimation, charging demand forecasting, and price prediction, though explainability features (confidence scores, rationale) are rarely mentioned.
    \item Market Priorities: Fleet operators emphasize uptime, while \gls{cpms} vendors focus on cost optimisation. Closed-loop operations, full automated decision making without human-in-the-loop, are widely regarded as an ideal but remain far from reality due to slow technology adoption and organizational barriers.
\end{enumerate}

\section{Discussion}

\subsection{Limitations}
This review provides a snapshot of current practices in digital systems for operational decision making in electric fleet management. However, several limitations must be acknowledged. First, the reliance on publicly available information introduces inherent incompleteness; proprietary features and functionalities protected by intellectual property rights are often not disclosed. Second, the search process is subject to location-based bias and search engine personalization, despite efforts to mitigate these effects by varying IP locations and disabling personalization settings. Third, the grey literature approach cannot guarantee exhaustiveness—while stopping criteria were applied, emerging solutions may have been overlooked. Fourth, larger companies which offer several different software systems might offer specific features independently without having them integrated in a single software system, but this is not always clear from their publicly available information. Fifth, details on implementation are typically not given, which means that a company mentioning they have implemented smart-charging might have a smart-charging function that does not produce good results. Finally, potential biases from both researchers and respondents during interviews may influence interpretation of trends.

\subsection{Industry impact}
Despite these limitations, the findings highlight a clear trajectory toward increased automation and integration across operational decision layers—site energy management, fleet scheduling, and charging session coordination. Industry actors are progressively embedding advanced functionalities such as dynamic load management, real-time telematics integration, and predictive algorithms for route and charging optimisation. These developments reflect the growing complexity introduced by electrification and the need for robust, data-driven solutions. However, adoption remains uneven: while charging session management shows significant innovation, site-level energy decision-making often lacks fleet-specific integration, and interoperability between systems remains a critical gap. Addressing these integration challenges will be essential for unlocking the full potential of smart charging and cost optimisation in large-scale fleet operations. Electrification sharply increases the data planners should process, from telematics and fleet schedules to site energy constraints and charger availability. This rising cognitive load, combined with fragmented and partially observable systems, makes manual oversight increasingly unreliable or less optimal. As a result, higher levels of automation and decision support are becoming essential as BEV fleets scale. However, it must also be acknowledged that traditional logistics operations today are not fully optimal, and some foundational issues in current planning workflows still need to be solved.

\section{Summary and recommendations}
This review highlights insights into the current state of digital systems for operational decision-making in electric fleet logistics.\par
First, fleet scheduling solutions exhibit the lowest level of adoption and maturity, largely due to the limited penetration of \glspl{bev} and conservative planning practices. Many operators continue to rely on manual or spreadsheet-based tools, which lack the sophistication required to handle the complexities introduced by electrification.\par
Second, co-optimisation across site-level energy management, charging session management, and fleet scheduling remains absent in practice. Current systems operate in silos, exchanging data through one-way APIs without unified optimisation engines. This fragmentation leads to partially observable decision-making and potential suboptimal outcomes, particularly when site constraints and charging priorities conflict. \par
Third, there are significant maturity differences between the three operational layers. \gls{cpms} platforms are the most advanced, offering dynamic load management and emerging predictive features. \gls{ems} solutions are well-developed for traditional energy tasks but slow to incorporate fleet-specific capabilities. Fleet scheduling tools lag far behind, with most EV-aware features still on vendor roadmaps. \par

\subsection{Answering the research questions}
\begin{questions}
    \item \textit{Which operational decisions related to electric fleet management are currently automated and which decisions remain manual or unsupported?}\par
    Automation is concentrated in charging management (CPMS), where functions such as load balancing, prioritisation, and tariff-based optimisation are widely implemented. Site energy systems automate traditional energy decisions (e.g., BESS scheduling, grid constraint management) but rarely incorporate fleet inputs. In contrast, fleet scheduling remains largely manual or planner-in-the-loop, with only limited EV-specific functionality. End-to-end automated decision-making across layers is not observed.
    \item \textit{Which industry actors provide solutions for which kind of automated decision making, what functionalities do they offer and what do these systems optimise for?}
    CPMS providers optimise charging operations for cost, utilisation, and power constraints, while EMS providers focus on site-level energy efficiency and grid interaction. Fleet management software prioritises operational feasibility, service levels, and routing constraints. These systems optimise within their respective domains, with limited alignment of objectives across actors. As a result, optimisation remains fragmented.
    \item \textit{To what extent do existing solutions enable automation and integration—ranging from decision support to fully optimised operational plans?}
    Integration between systems is limited and primarily consists of one-way data exchange via APIs and protocols such as OCPP. Bi-directional coordination and joint optimisation across EMS, CPMS, and fleet scheduling are largely absent. Automation is therefore confined to individual systems, resulting in partially observable decision-making. Fully integrated, closed-loop operational planning is not yet reflected in current practice.
    \item \textit{What types of data and information sources (e.g., vehicle telematics, energy prices, grid constraints) are incorporated into these solutions for decision-making?}
    Key data inputs include vehicle telematics (e.g., state of charge), electricity prices, site power availability, and charger status. EMS systems additionally rely on generation and storage data, while fleet tools incorporate routing and traffic information. Although these data sources are individually available, their cross-system integration remains limited. This constrains their use for coordinated optimisation.
\end{questions}

\subsection{Future research priorities}
Interoperability frameworks and algorithms for multi-layer co-optimisation represent the most promising directions. These solutions must jointly consider site energy constraints, charging priorities, and fleet scheduling objectives under uncertainty. Research should show which improved utilisation of assets and cost saving can potentially be generated through better coordination between decision making systems, or how much performance is lost because of compartmentalisation. Some work in this direction has already been done by \citet{tno_optimal_charging_logistics_2024}. Developing standardized communication protocols between EMS, CPMS, and fleet management systems—supporting not only data exchange but also coordinated decision-making—remains an important research direction. 
Finally, while this paper does not dedicate a section to Industry 4.0 \citep{schwab2017fourth}, it is worth noting that digitalization trends—such as IoT connectivity, AI-driven predictive analytics, and digital twins—will play a supportive role in enabling the electrification of logistics fleets \citep{liu2019internet}. These technologies should be integrated pragmatically, complementing rather than replacing foundational improvements in interoperability and optimisation.

\section*{Acknowledgements}
This work has been financially supported by Shell Global Solution International B.V. The opinions, findings, and conclusions, expressed in this publication, are those of the authors and do not necessarily reflect the views of Shell Global Solutions International B.V. \par
A market scan on fleet management software addressing the needs of electric vehicles has been executed by Evalueserve AG, which has been used as input for this paper. 
\bibliographystyle{elsarticle-num-names} 
\bibliography{biobliography}

@article{adaikkappan2022modeling,
  title={Modeling, state of charge estimation, and charging of lithium-ion battery in electric vehicle: a review},
  author={Adaikkappan, Maheshwari and Sathiyamoorthy, Nageswari},
  journal={International Journal of Energy Research},
  volume={46},
  number={3},
  pages={2141--2165},
  year={2022},
  publisher={Wiley Online Library}
}

@article{gur2018review,
  title={Review of electrical energy storage technologies, materials and systems: challenges and prospects for large-scale grid storage},
  author={G{\"u}r, Turgut M},
  journal={Energy \& Environmental Science},
  volume={11},
  number={10},
  pages={2696--2767},
  year={2018},
  publisher={Royal Society of Chemistry}
}

@article{liu2019internet,
  title={An ‘Internet of Things’ enabled dynamic optimization method for smart vehicles and logistics tasks},
  author={Liu, Sichao and Zhang, Yingfeng and Liu, Yang and Wang, Lihui and Wang, Xi Vincent},
  journal={Journal of cleaner production},
  volume={215},
  pages={806--820},
  year={2019},
  publisher={Elsevier}
}

@book{schwab2017fourth,
  title={The fourth industrial revolution},
  author={Schwab, Klaus},
  year={2017},
  publisher={Crown Currency}
}

@article{bloomberg2024,
  title={Electric Vehicle Outlook 2024},
  author={BloombergNEF},
  journal={BloombergNEF},
  year={2024}
}

@article{verma2022life,
  title={Life cycle assessment of electric vehicles in comparison to combustion engine vehicles: A review},
  author={Verma, Shrey and Dwivedi, Gaurav and Verma, Puneet},
  journal={Materials Today: Proceedings},
  volume={49},
  pages={217--222},
  year={2022},
  publisher={Elsevier}
}

@inproceedings{kliman2015location,
  title={Location, location, location: The impact of geolocation on web search personalization},
  author={Kliman-Silver, Chloe and Hannak, Aniko and Lazer, David and Wilson, Christo and Mislove, Alan},
  booktitle={Proceedings of the 2015 internet measurement conference},
  pages={121--127},
  year={2015}
}

@article{garousi2019guidelines,
  title={Guidelines for including grey literature and conducting multivocal literature reviews in software engineering},
  author={Garousi, Vahid and Felderer, Michael and M{\"a}ntyl{\"a}, Mika V},
  journal={Information and software technology},
  volume={106},
  pages={101--121},
  year={2019},
  publisher={Elsevier}
}

@techreport{kirby2003frequency,
  title={Frequency control concerns in the North American electric power system},
  author={Kirby, Brendan J},
  year={2003},
  institution={Oak Ridge National Lab.(ORNL), Oak Ridge, TN (United States)}
}

@misc{tennet2024marketreport,
title={Annual Market Update 2024: Electricity Market Insights},
author={TenneT},
year={2025},
note={Accessed September 2, 2025}
}

@misc{paris2015,  
    title   = {Paris Agreement},
    author  = {{United Nations Framework Convention on Climate Change}}, 
    year    = {2015},  
    note    = {Adopted at COP21, Paris},  
    url     = {https://unfccc.int/sites/default/files/english_paris_agreement.pdf}
}

@misc{daimler_eactros_2021,  
    author = {Daimler},  
    title = {A new truck for a new era: Mercedes-Benz eActros celebrates its world premiere},  
    year = {2021},  
    url = {https://www.daimlertruck.com/en/newsroom/pressrelease/a-new-truck-for-a-new-era-mercedes-benz-eactros-celebrates-its-world-premiere-50352780},  
    note = {Series production scheduled to begin in H2 2021 at the Wörth plant}
}

@misc{volvo2019electric,  
    author       = {Volvo},  
    title        = {Volvo Trucks starts series production of electric trucks},  
    year         = {2019},  
    month        = {November},  
    url          = {https://www.volvotrucks.com/en-en/news-stories/press-releases/2019/nov/pressrelease-191106.html},  
    note         = {Accessed: 2025-11-03}
}

@misc{renault2023etech,  
    author       = {Renault},  
    title        = {Series production starts in Bourg-en-Bresse for Renault Trucks E-Tech T and C},  
    year         = {2023},  
    month        = {November},  
    url          = {https://www.renault-trucks.com/en/newsroom/press-releases/series-production-starts-bourg-en-bresse-renault-trucks-e-tech-t-c},  
    note         = {Accessed: 2025-11-03}
}

@misc{man2025electric,  
    author       = {MAN},  
    title        = {Historic: MAN starts series production of electric trucks},  
    year         = {2025},  
    month        = {June},  
    url          = {https://press.mantruckandbus.com/corporate/historic-man-starts-series-production-of-electric-trucks/},  
    note         = {Accessed: 2025-11-03}
}

@misc{daf2023electricplant,  
    author       = {DAF},  
    title        = {New DAF Electric Truck Assembly Plant Officially Opened},  
    year         = {2023},  
    month        = {April},    
    url          = {https://www.daf.global/en-us/news-and-media/news-articles/global/2023/24-04-2023-new-daf-electric-truck-assembly-plant-officially-opened},  
    note         = {Accessed: 2025-11-03}
}

@misc{iveco2023pressrelease,  
    author       = {IVECO},  
    title        = {IVECO to produce and market its Heavy-Duty Battery Electric Vehicle and Heavy-Duty Fuel Cell Electric Vehicle},  
    year         = {2023},  
    month        = {November},  
    url          = {https://www.iveco.com/global/Press/PressReleases/2023/IVECO-to-produce-and-market-its-Heavy-Duty-Battery-Electric-Vehicle-and-Heavy-Duty-Fuel-Cell-Electric-Vehicle},  
    note         = {Accessed: 2025-11-03}
}

@misc{scania2020electriclaunch,  
    author       = {Scania},  
    title        = {2020: First Battery Electric Truck Launch},  
    year         = {2020},  
    url          = {https://www.scania.com/group/en/home/about-scania/heritage/corporate-milestones/2020-first-battery-electric-truck-launch.html},  
    note         = {Accessed: 2025-11-03}
}

@article{clo2015merit,
  title={The merit-order effect in the Italian power market: The impact of solar and wind generation on national wholesale electricity prices},
  author={Cl{\`o}, Stefano and Cataldi, Alessandra and Zoppoli, Pietro},
  journal={Energy Policy},
  volume={77},
  pages={79--88},
  year={2015},
  publisher={Elsevier}
}

@article{pavlik2025renewable,
  title={Renewable Energy and Price Stability: An Analysis of Volatility and Market Shifts in the European Electricity Sector (2015--2025).},
  author={Pavl{\'\i}k, Marek and Kurimsk{\`y}, Franti{\v{s}}ek and {\v{S}}evc, Kamil},
  journal={Applied Sciences (2076-3417)},
  volume={15},
  number={12},
  year={2025}
}

@misc{CBS_80416ENG_misc,
  title        = {Pump prices motor fuels; per type, per day},
  author       = {{Statistics Netherlands (CBS)}},
  howpublished = {CBS StatLine},
  year         = {2025},
  url          = {https://www.cbs.nl/en-gb/figures/detail/80416ENG},
  note         = {Table code: 80416ENG; Source: Travelcard BV; Data available from January 2006; Figures are final. Accessed 2025-12-24}
}

@misc{EC_DGENER_WeeklyOilBulletin_misc,
  title        = {Weekly Oil Bulletin},
  author       = {{European Commission, Directorate-General for Energy}},
  howpublished = {European Commission --- DG Energy, Data and Analysis},
  year         = {2025},
  url          = {https://energy.ec.europa.eu/data-and-analysis/weekly-oil-bulletin_en},
  note         = {Accessed 2025-12-24; publisher: European Commission (DG Energy); see dataset record on data.europa.eu},
}

@article{shahzad2023possibilities,
  title={Possibilities, challenges, and future opportunities of microgrids: A review},
  author={Shahzad, Sulman and Abbasi, Muhammad Abbas and Ali, Hassan and Iqbal, Muhammad and Munir, Rania and Kilic, Heybet},
  journal={Sustainability},
  volume={15},
  number={8},
  pages={6366},
  year={2023},
  publisher={MDPI}
}

@article{salman2022review,
  title={A review of improvements in power system flexibility: implementation, operation and economics},
  author={Salman, Umar Taiwo and Shafiq, Saifullah and Al-Ismail, Fahad S and Khalid, Muhammad},
  journal={Electronics},
  volume={11},
  number={4},
  pages={581},
  year={2022},
  publisher={MDPI}
}

@online{CBS_2025_RenewablesHalf,  title     = {Half of electricity is produced from renewable sources},  author    = {{Statistics Netherlands (CBS)}},  year      = {2025},  url       = {https://www.cbs.nl/en-gb/news/2025/11/half-of-electricity-is-produced-from-renewable-sources},  urldate   = {2025-12-24},  note      = {Official national statistics; reports \~50\% renewable electricity in 2024, peak 63\% in April}}

@report{IEA_GlobalEVOutlook_2025,
  title        = {Global EV Outlook 2025: Expanding sales in diverse markets},
  author       = {{International Energy Agency (IEA)}},
  institution  = {International Energy Agency},
  year         = {2025},
  url          = {https://www.iea.org/reports/global-ev-outlook-2025},
  urldate      = {2025-12-24},
  note         = {Licence: CC BY 4.0; PDF: https://iea.blob}
}

@manual{Elaflex_ZVA25_HiFlo_Datasheet_2008,  title        = {ZVA 25 HiFlo Automatic Nozzle (1): Datasheet},  author       = {{ELAFLEX HIBY GmbH \& Co. KG}},  year         = {2008},  address      = {Hamburg, Germany},  note         = {Datasheet for ZVA 25 (DN\,25) HiFlo nozzle. Flow rate up to 140\,L/min at 0.5–3.5\,bar; variants and accessories listed; EN~13012/ATEX/weights \& measures},  url          = {https://commercialfuelsolutions.co.uk/downloads/data_sheets/ZVA25.pdf},  urldate      = {2025-12-24}}

@inproceedings{RefaEVS38_2025,
  author       = {Refa, Nazir and Janssen, Jeroen and van der Geer, Gijsbert and Broos, Paul and Bos, Thomas},
  title        = {Projecting the Spatial Distribution of Charging Infrastructure Demand for Battery-Electric Trucks and Vans in the Netherlands},
  booktitle    = {Proceedings of the 38th International Electric Vehicle Symposium and Exhibition (EVS38)},
  address      = {Gothenburg, Sweden},
  date         = {2025-06-15/2025-06-18},
  year         = {2025},
  url          = {https://evs38-program.org/images/Proceedings/D%20Charging%20Infrastructure%20and%20grid%20integration/421_Projecting%20the%20Spatial%20Distribution%20of%20charging%20Infrastructure%20Demand%20for%20battery-Electric%20Trucks%20and%20Vans%20in%20the%20Netherlands.pdf},
  urldate      = {2025-12-24},
  note         = {ElaadNL paper; EVS38 session: Charging Infrastructure and Grid Integration}
}

@report{Hacker_Truck_Depot_Charging_2025,
  author      = {Hacker, F. and Gnann, T. and Le Corguill{\'e}, J. and Stephan, A. and Kappler, L. and Mottschall, M. and Pl{\"o}tz, P.},
  title       = {Truck depot charging},
  date        = {2025},
  year        = {2025},
  institution = {Fraunhofer ISI; {\"O}ko-Institut},
  location    = {Karlsruhe; Berlin},
  note        = {Study commissioned by Transport \& Environment}
}

@online{Gartner_TMS_PeerInsights_2025,  
    title           ={Best Transportation Management Systems Reviews 2025 | Gartner Peer Insights},
    year            = {2025},
    url             = {https://www.gartner.com/reviews/market/transportation-management-systems},  
    urldate         = {2025-12-24},  
    organization    = {Gartner Peer Insights},  
    author          = {Gartner, Inc.},  
    language        = {en},  
    note            = {Market category page with user reviews and vendor listings}}

@online{VDV463_ShortIntro_2025,  author       = {Dohmen, Claus},  title        = {VDV463: Charging management interface for electric busses},  year         = {2025},  month        = {6},  address      = {Hamburg, Germany},  organization = {Verband Deutscher Verkehrsunternehmen (VDV)},  url          = {https://www.vdv.de/v6-vdv463-charging-management-interface-v1.pdfx},  urldate      = {2025-12-24},  language     = {en},  note         = {Short introduction (slides) on the VDV463 interface; version history and links to the JSON schema provided}}

@techreport{ICCT2023HDVDecarb,  title        = {Decarbonizing Heavy-Duty Vehicles in Europe: Cost Analysis},  author       = {Rodríguez, César and Delgado, Jorge and Muncrief, Richard},  institution  = {International Council on Clean Transportation},  year         = {2023},  month        = {January},  url          = {https://theicct.org/wp-content/uploads/2023/01/hdv-europe-decarb-costs-jan23.pdf},  note         = {Accessed: 2026-01-09}}

@online{CharIN_MCS,  title        = {Megawatt Charging System (MCS)},  author       = {{CharIN e.V.}},  year         = {2026},  url          = {https://www.charin.global/technology/mcs/},  note         = {Accessed: 2026-01-09}}

@article{gray2021decarbonising,
  title={Decarbonising ships, planes and trucks: An analysis of suitable low-carbon fuels for the maritime, aviation and haulage sectors},
  author={Gray, Nathan and McDonagh, Shane and O'Shea, Richard and Smyth, Beatrice and Murphy, Jerry D},
  journal={Advances in Applied Energy},
  volume={1},
  pages={100008},
  year={2021},
  publisher={Elsevier}
}

@online{Plant_Fleetio_EV_Fleet_Management_2024,
  author       = {Plant, Rachael},
  title        = {Electric Vehicle Fleet Management Guide},
  organization = {Fleetio},
  date         = {2024-04-18},
  url          = {https://www.fleetio.com/blog/electric-vehicle-fleet-management},
  urldate      = {2026-01-13},
  note         = {Fleet Management Blog}
}

@misc{Driivz2026,
  author = {Driivz},
  title = {Driivz},
  year = {2026},
  howpublished = {\url{https://driivz.com/}},
  note = {Accessed January 07, 2026}
}

@misc{Ampeco2026,
  author = {Ampeco},
  title = {Ampeco},
  year = {2026},
  howpublished = {\url{https://www.ampeco.com/}},
  note = {Accessed January 07, 2026}
}

@misc{Monta2026,
  author = {Monta},
  title = {Monta},
  year = {2026},
  howpublished = {\url{https://monta.com/en/}},
  note = {Accessed January 07, 2026}
}

@misc{OceanEV2026,
  author = {Ocean EV},
  title = {Ocean EV Charging Platform},
  year = {2026},
  howpublished = {\url{https://oceanevcharging.com/}},
  note = {Accessed January 07, 2026}
}

@misc{Greenflux2026,
  author = {Greenflux},
  title = {Greenflux},
  year = {2026},
  howpublished = {\url{https://www.greenflux.com/}},
  note = {Accessed January 07, 2026}
}

@misc{LastMileSolution2026,
  author = {LastMileSolution},
  title = {LastMileSolution},
  year = {2026},
  howpublished = {\url{https://www.lastmilesolutions.com/smart-}},
  note = {Accessed January 07, 2026}
}

@misc{Vitra2026,
  author = {Vitra},
  title = {Vitra},
  year = {2026},
  howpublished = {\url{https://www.virta.global/}},
  note = {Accessed January 07, 2026}
}

@misc{Current2026,
  author = {Current},
  title = {Current},
  year = {2026},
  howpublished = {\url{https://www.current.eco/about-us}},
  note = {Accessed January 07, 2026}
}

@misc{SwitchEV2026,
  author = {SwitchEV},
  title = {SwitchEV},
  year = {2026},
  howpublished = {\url{https://www.switch-ev.com/solutions/}},
  note = {Accessed January 07, 2026}
}

@misc{Emabler2026,
  author = {Emabler},
  title = {Emabler},
  year = {2026},
  howpublished = {\url{https://emabler.com/}},
  note = {Accessed January 07, 2026}
}

@misc{ChargePoint2026,
  author = {ChargePoint},
  title = {ChargePoint},
  year = {2026},
  howpublished = {\url{https://www.chargepoint.com/fleet/software}},
  note = {Accessed January 07, 2026}
}

@misc{EOCharging2026,
  author = {EOCharging},
  title = {EOCharging},
  year = {2026},
  howpublished = {\url{https://www.eocharging.com/energy-}},
  note = {Accessed January 07, 2026}
}

@misc{ChargeLab2026,
  author = {ChargeLab},
  title = {ChargeLab},
  year = {2026},
  howpublished = {\url{https://chargelab.co/software}},
  note = {Accessed January 07, 2026}
}

@misc{EVConnect2026,
  author = {EVConnect},
  title = {EVConnect},
  year = {2026},
  howpublished = {\url{https://www.evconnect.com/services/}},
  note = {Accessed January 07, 2026}
}

@misc{TridensTechnology2026,
  author = {Tridens Technology},
  title = {Tridens Technology},
  year = {2026},
  howpublished = {\url{https://tridenstechnology.com/}},
  note = {Accessed January 07, 2026}
}

@misc{ChargePanel2026,
  author = {ChargePanel},
  title = {ChargePanel},
  year = {2026},
  howpublished = {\url{https://www.chargepanel.com/enterprise-}},
  note = {Accessed January 07, 2026}
}

@misc{Reev2026,
  author = {Reev},
  title = {Reev},
  year = {2026},
  howpublished = {\url{https://reev.com/software}},
  note = {Accessed January 07, 2026}
}

@misc{Evoltsoft2026,
  author = {Evoltsoft},
  title = {Evoltsoft},
  year = {2026},
  howpublished = {\url{https://www.evoltsoft.com/}},
  note = {Accessed January 07, 2026}
}

@misc{ClenergyEV2026,
  author = {ClenergyEV},
  title = {ClenergyEV},
  year = {2026},
  howpublished = {\url{https://www.clenergy-ev.com/}},
  note = {Accessed January 07, 2026}
}

@misc{Vialumina2026,
  author = {Vialumina},
  title = {Vialumina},
  year = {2026},
  howpublished = {\url{https://vialumina.ai/solution}},
  note = {Accessed January 07, 2026}
}

@misc{Deftpower2026,
  author = {Deftpower},
  title = {Deftpower},
  year = {2026},
  howpublished = {\url{https://www.deftpower.com}},
  note = {Accessed January 07, 2026}
}

@misc{Plugsurfing2026,
  author = {Plugsurfing},
  title = {Plugsurfing},
  year = {2026},
  howpublished = {\url{https://www.plugsurfing.com}},
  note = {Accessed January 07, 2026}
}

@misc{Bluedot2026,
  author = {Bluedot},
  title = {Bluedot},
  year = {2026},
  howpublished = {\url{https://www.bluedot.co/}},
  note = {Accessed January 07, 2026}
}

@online{nuvve_platform,  title   = {Nuvve: Vehicle-to-Grid (V2G) and EV Charging Solutions},  author  = {Nuvve},  year    = {2026},  url     = {https://nuvve.com/},  note    = {Official website of Nuvve, detailing V2G technology, fleet solutions, and energy management services},  urldate = {2026-01-05}}

@misc{eaton_brightlayer_energy,  title        = {Brightlayer Energy Management Software},  author       = {Eaton},  year         = {2026},  url          = {https://www.eaton.com/us/en-us/catalog/software/brightlayer-energy.html},  note         = {Accessed: 2026-01-10}}

@misc{Spirii2026,
  author = {Spirii},
  title = {Spirii},
  year = {2026},
  howpublished = {\url{https://www.spirii.com/en}},
  note = {Accessed January 07, 2026}
}

@misc{HitachiZeroCarbon2026,
  author = {Hitachi Zero Carbon},
  title = {Hitachi Zero Carbon},
  year = {2026},
  howpublished = {\url{https://www.hitachizerocarbon.com/news-}},
  note = {Accessed January 07, 2026}
}

@misc{RiDERgy2026,
  author = {RiDERgy},
  title = {RiDERgy},
  year = {2026},
  howpublished = {\url{https://www.ridergy.com/services}},
  note = {Accessed January 07, 2026}
}

@misc{Synop2026,
  author = {Synop},
  title = {Synop},
  year = {2026},
  howpublished = {\url{https://www.synop.ai/}},
  note = {Accessed January 07, 2026}
}

@misc{PowerON2026,
  author = {PowerON},
  title = {PowerON},
  year = {2026},
  howpublished = {\url{https://poweronenergy.ca/solutions/}},
  note = {Accessed January 07, 2026}
}

@misc{FutureEnergy2026,
  author = {FutureEnergy},
  title = {FutureEnergy},
  year = {2026},
  howpublished = {\url{https://futureenergy.com/solutions/smart-}},
  note = {Accessed January 07, 2026}
}

@misc{BluwaveAI2026,
  author = {Bluwave-AI},
  title = {Bluwave-AI},
  year = {2026},
  howpublished = {\url{https://www.bluwave-ai.com/ev-fleet-}},
  note = {Accessed January 07, 2026}
}

@misc{ABB2026,
  author = {ABB},
  title = {ABB},
  year = {2026},
  howpublished = {\url{https://new.abb.com/process-automation/}},
  note = {Accessed January 07, 2026}
}

@misc{Ampcontrol2026,
  author = {Ampcontrol},
  title = {Ampcontrol},
  year = {2026},
  howpublished = {\url{https://www.ampcontrol.io/nl}},
  note = {Accessed January 07, 2026}
}

@misc{Ampowr2026,
  author = {Ampowr},
  title = {Ampowr},
  year = {2026},
  howpublished = {\url{https://ampowr.com/industries/ev-}},
  note = {Accessed January 07, 2026}
}

@misc{Honeywell2026,
  author = {Honeywell},
  title = {Honeywell},
  year = {2026},
  howpublished = {\url{https://buildings.honeywell.com/us/en/lp/}},
  note = {Accessed January 07, 2026}
}

@misc{iWell2026,
  author = {iWell},
  title = {iWell},
  year = {2026},
  howpublished = {\url{https://iwell.eu/product-overview}},
  note = {Accessed January 07, 2026}
}

@misc{Schneider2026,
  author = {Schneider},
  title = {Schneider},
  year = {2026},
  howpublished = {\url{https://www.se.com/sg/en/work/campaign/}},
  note = {Accessed January 07, 2026}
}

@misc{Ampfox2026,
  author = {Ampfox},
  title = {Ampfox},
  year = {2026},
  howpublished = {\url{https://www.ampfox.co/}},
  note = {Accessed January 07, 2026}
}

@misc{MichelinConnectedFleet2026,
  author = {Michelin Connected Fleet},
  title = {Michelin Connected Fleet},
  year = {2026},
  howpublished = {\url{https://connectedfleet.michelin.com/}},
  note = {Accessed January 07, 2026}
}

@misc{LocoNav2026,
  author = {LocoNav},
  title = {LocoNav},
  year = {2026},
  howpublished = {\url{https://loconav.com/}},
  note = {Accessed January 07, 2026}
}

@misc{TrakM82026,
  author = {TrakM8},
  title = {TrakM8},
  year = {2026},
  howpublished = {\url{https://www.trakm8.com/fleet-management/ev-}},
  note = {Accessed January 07, 2026}
}

@misc{KiaPBVFMS2026,
  author = {Kia PBV FMS},
  title = {Kia PBV FMS},
  year = {2026},
  howpublished = {\url{https://www.kia.com/eu/pbv/home/}},
  note = {Accessed January 07, 2026}
}

@misc{Trimble2026,
  author = {Trimble},
  title = {Trimble},
  year = {2026},
  howpublished = {\url{https://transportation.trimble.com/}},
  note = {Accessed January 07, 2026}
}

@misc{GPSInsight2026,
  author = {GPS Insight},
  title = {GPS Insight},
  year = {2026},
  howpublished = {\url{https://www.gpsinsight.com/field-service/}},
  note = {Accessed January 07, 2026}
}

@misc{XotrucksXosphere2026,
  author = {Xotrucks/Xosphere},
  title = {Xotrucks/Xosphere},
  year = {2026},
  howpublished = {\url{https://www.xostrucks.com/xosphere}},
  note = {Accessed January 07, 2026}
}

@misc{FYND2026,
  author = {FYND},
  title = {FYND},
  year = {2026},
  howpublished = {\url{https://www.fynd.com/solutions/transport-}},
  note = {Accessed January 07, 2026}
}

@misc{SmartTrak2026,
  author = {SmartTrak},
  title = {SmartTrak},
  year = {2026},
  howpublished = {\url{https://smartrak.com/electric-vehicle-}},
  note = {Accessed January 07, 2026}
}

@misc{GridLine2026,
  author = {GridLine},
  title = {GridLine},
  year = {2026},
  howpublished = {\url{https://www.gridline.com/fleet-management-}},
  note = {Accessed January 07, 2026}
}

@misc{Volteras2026,
  author = {Volteras},
  title = {Volteras},
  year = {2026},
  howpublished = {\url{https://www.volteras.com/solutions/route-}},
  note = {Accessed January 07, 2026}
}

@misc{ZeroMission2026,
  author = {Zero Mission},
  title = {Zero Mission},
  year = {2026},
  howpublished = {\url{https://zeromission.io/operate-platform}},
  note = {Accessed January 07, 2026}
}

@misc{RoadCast2026,
  author = {RoadCast},
  title = {RoadCast},
  year = {2026},
  howpublished = {\url{https://roadcast.in/blog/intelligent-}},
  note = {Accessed January 07, 2026}
}

@misc{Fryte2026,
  author = {Fryte},
  title = {Fryte},
  year = {2026},
  howpublished = {\url{https://www.fryte.eu/en}},
  note = {Accessed January 07, 2026}
}

@misc{eMotionFleet2026,
  author = {eMotionFleet},
  title = {eMotionFleet},
  year = {2026},
  howpublished = {\url{https://www.emotion-fleet.com/en/}},
  note = {Accessed January 07, 2026}
}

@misc{TeletracNavman2026,
  author = {Teletrac Navman},
  title = {Teletrac Navman},
  year = {2026},
  howpublished = {\url{https://www.teletracnavman.com/fleet-}},
  note = {Accessed January 07, 2026}
}

@misc{BetterFleet2026,
  author = {BetterFleet},
  title = {BetterFleet},
  year = {2026},
  howpublished = {\url{https://www.betterfleet.com/}},
  note = {Accessed January 07, 2026}
}

@misc{OptiBus2026,
  author = {OptiBus},
  title = {OptiBus},
  year = {2026},
  howpublished = {\url{https://optibus.com/}},
  note = {Accessed January 07, 2026}
}

@misc{MoevAI2026,
  author = {Moev.AI},
  title = {Moev.AI},
  year = {2026},
  howpublished = {\url{https://www.moev.ai/}},
  note = {Accessed January 07, 2026}
}

@misc{Tenix2026,
  author = {Tenix},
  title = {Tenix},
  year = {2026},
  howpublished = {\url{https://tenix.eu/fleet/}},
  note = {Accessed January 07, 2026}
}

@misc{Katsana2026,
  author = {Katsana},
  title = {Katsana},
  year = {2026},
  howpublished = {\url{https://www.katsana.com/ev/}},
  note = {Accessed January 07, 2026}
}

@misc{Chetu2026,
  author = {Chetu},
  title = {Chetu},
  year = {2026},
  howpublished = {\url{https://www.chetu.com/transportation/}},
  note = {Accessed January 07, 2026}
}

@misc{ZonarSystems2026,
  author = {Zonar Systems},
  title = {Zonar Systems},
  year = {2026},
  howpublished = {\url{https://www.zonarsystems.com/solutions/}},
  note = {Accessed January 07, 2026}
}

@misc{MagentaMobility2026,
  author = {Magenta Mobility},
  title = {Magenta Mobility},
  year = {2026},
  howpublished = {\url{https://www.magentamobility.com/}},
  note = {Accessed January 07, 2026}
}

@misc{MakeMyDay2026,
  author = {MakeMyDay},
  title = {MakeMyDay},
  year = {2026},
  howpublished = {\url{https://www.makemydayapp.com/}},
  note = {Accessed January 07, 2026}
}

@misc{AutoWiz2026,
  author = {AutoWiz},
  title = {AutoWiz},
  year = {2026},
  howpublished = {\url{https://www.autowiz.in/ev.html}},
  note = {Accessed January 07, 2026}
}

@misc{FleetX2026,
  author = {FleetX},
  title = {FleetX},
  year = {2026},
  howpublished = {\url{https://www.fleetx.io/electric-vehicles}},
  note = {Accessed January 07, 2026}
}

@misc{TrinetraWireless2026,
  author = {Trinetra Wireless},
  title = {Trinetra Wireless},
  year = {2026},
  howpublished = {\url{https://www.trinetrawireless.com/electric-}},
  note = {Accessed January 07, 2026}
}

@misc{Trackobit2026,
  author = {Trackobit},
  title = {Trackobit},
  year = {2026},
  howpublished = {\url{https://trackobit.com/electric-vehicle-}},
  note = {Accessed January 07, 2026}
}

@misc{Loginext2026,
  author = {Loginext},
  title = {Loginext},
  year = {2026},
  howpublished = {\url{https://www.loginextsolutions.com/blog/}},
  note = {Accessed January 07, 2026}
}

@misc{GoMotive2026,
  author = {GoMotive},
  title = {GoMotive},
  year = {2026},
  howpublished = {\url{https://gomotive.com/blog/ev-fleet-}},
  note = {Accessed January 07, 2026}
}

@misc{FarEye2026,
  author = {FarEye},
  title = {FarEye},
  year = {2026},
  howpublished = {\url{https://fareye.com/solutions/ev-route-}},
  note = {Accessed January 07, 2026}
}

@misc{NextBillionAI2026,
  author = {NextBillion.AI},
  title = {NextBillion.AI},
  year = {2026},
  howpublished = {\url{https://nextbillion.ai/use-cases/ev-route-}},
  note = {Accessed January 07, 2026}
}

@misc{Iveco2026,
  author = {Iveco},
  title = {Iveco},
  year = {2026},
  howpublished = {\url{https://www.iveco.com/uk/Services/e-}},
  note = {Accessed January 07, 2026}
}

@misc{PTVLogistics2026,
  author = {{PTV Logistics}},
  title = {PTV Logistics},
  year = {2026},
  howpublished = {\url{https://www.ptvlogistics.com/en/products/}},
  note = {Accessed January 07, 2026}
}

@misc{Ortec2026,
  author = {Ortec},
  title = {Ortec},
  year = {2026},
  howpublished = {\url{https://ortec.com/en/insights/e-mobility-}},
  note = {Accessed January 07, 2026}
}

@misc{Navixy2026,
  author = {Navixy},
  title = {Navixy},
  year = {2026},
  howpublished = {\url{https://www.navixy.com/}},
  note = {Accessed January 07, 2026}
}

@misc{StandardFleet2026,
  author = {StandardFleet},
  title = {StandardFleet},
  year = {2026},
  howpublished = {\url{https://www.standardfleet.com/}},
  note = {Accessed January 07, 2026}
}

@misc{MatrackInc2026,
  author = {{Matrack Inc}},
  title = {Matrack Inc},
  year = {2026},
  howpublished = {\url{https://matrackinc.com/}},
  note = {Accessed January 07, 2026}
}

@misc{FleetBoard2026,
  author = {FleetBoard},
  title = {FleetBoard},
  year = {2026},
  howpublished = {\url{https://www.fleetboard.info/digital-}},
  note = {Accessed January 07, 2026}
}

@misc{ConsatTelematics2026,
  author = {{Consat Telematics}},
  title = {Consat Telematics},
  year = {2026},
  howpublished = {\url{https://www.consat.com/telematics/}},
  note = {Accessed January 07, 2026}
}

@misc{Webfleet2026,
  author = {Webfleet},
  title = {Webfleet},
  year = {2026},
  howpublished = {\url{https://www.webfleet.com/en_gb/webfleet/}},
  note = {Accessed January 07, 2026}
}

@misc{BlueArrowTelematics2026,
  author = {{BlueArrow Telematics}},
  title = {BlueArrow Telematics},
  year = {2026},
  howpublished = {\url{https://bluearrowtelematics.com/fleet-}},
  note = {Accessed January 07, 2026}
}

@misc{TargaTelematics2026,
  author = {{Targa Telematics}},
  title = {Targa Telematics},
  year = {2026},
  howpublished = {\url{https://targatelematics.com/solutions/}},
  note = {Accessed January 07, 2026}
}

@misc{Geotab2026,
  author = {Geotab},
  title = {Geotab},
  year = {2026},
  howpublished = {\url{https://www.geotab.com/fleet-management-}},
  note = {Accessed January 07, 2026}
}

@misc{Scania2026,
  author = {Scania},
  title = {Scania},
  year = {2026},
  howpublished = {\url{https://www.scania.com/group/en/home/}},
  note = {Accessed January 07, 2026}
}

@misc{Ituran2026,
  author = {Ituran},
  title = {Ituran},
  year = {2026},
  howpublished = {\url{https://www.ituran.com/ituranglobal/}},
  note = {Accessed January 07, 2026}
}

@misc{Samsara2026,
  author = {Samsara},
  title = {Samsara},
  year = {2026},
  howpublished = {\url{https://www.samsara.com/products/}},
  note = {Accessed January 07, 2026}
}

@misc{Codibly2026,
  author = {Codibly},
  title = {Codibly},
  year = {2026},
  howpublished = {\url{https://codibly.com/e-mobility/ems-bess-}},
  note = {Accessed January 07, 2026}
}

@misc{Tekmindz2026,
  author = {Tekmindz},
  title = {Tekmindz},
  year = {2026},
  howpublished = {\url{https://www.tekmindz.com/industries/energy-}},
  note = {Accessed January 07, 2026}
}

@misc{ZealousSystem2026,
  author = {{Zealous System}},
  title = {Zealous System},
  year = {2026},
  howpublished = {\url{https://www.zealousys.com/blog/develop-}},
  note = {Accessed January 07, 2026}
}

@misc{Here2026,
  author = {Here},
  title = {Here},
  year = {2026},
  howpublished = {\url{https://www.here.com/solutions/ev-routing-}},
  note = {Accessed January 07, 2026}
}

@misc{chargetrip,
  title        = {ChargeTrip},
  author       = {{ChargeTrip}},
  year         = {2026},
  url          = {https://www.chargetrip.com/},
  note         = {Accessed: 26 January 2026}
}

@online{tno_optimal_charging_logistics_2024,
  title   = {Optimal charging planning for logistics service providers considering grid congestion},
  author  = {{TNO}},
  year    = {2024},
  month   = {dec},
  url     = {https://www.tno.nl/en/newsroom/insights/2024/12/optimal-charging-planning-logistics/},
  urldate = {2026-01-26}
}

@online{hubject_website,
  title        = {Hubject},
  author       = {Hubject},
  url          = {https://www.hubject.com/},
  urldate      = {2026-02-26}
}

@online{dhemax_website,
  title   = {Dhemax},
  author = {Dhemax},
  url     = {https://www.dhemax.com/en/},
  urldate = {2026-02-26}
}

@online{inceptev2026,
  title   = {InceptEV},
  author = {InceptEV},
  url     = {https://www.inceptev.com/},
  urldate = {2026-02-26}
}
\appendix
\onecolumn
\section{Findings tables}\label{sec:findings_tables}

\subsection{CPMS}
\begin{table}[H]
\resizebox{\textwidth}{!}{%
\begin{tabular}{|p{2cm}|p{1cm}|p{2.5cm}|p{2.5cm}|p{2.5cm}|p{3.5cm}|}
\hline
\textbf{Capability / Decision}                                          & \textbf{Market Prevalence}     & \textbf{How It Is Done Today}              & \textbf{Supporting Data (Up- and Downstream)}                           & \textbf{Observed Limitations / Risks}                             & \textbf{Reference Companies} \\ \hline
Dynamic load balancing                             & High                                            & Per-site cap; per-connector throttling            & Site metre; sub-metering                                           & Myopic if EMS setpoints static                           & \cite{Driivz2026, Ampeco2026, Monta2026, OceanEV2026, Greenflux2026, LastMileSolution2026, Vitra2026, Current2026, SwitchEV2026, Emabler2026, ChargePoint2026, EOCharging2026, ChargeLab2026, EVConnect2026, TridensTechnology2026, ChargePanel2026, Reev2026, Evoltsoft2026, ClenergyEV2026, Vialumina2026}                                          \\ \hline
Priority queues / charging policies                & High                                            & Rule-based priority per vehicle/connector         & Manual/UI policy; CSV                                              & No optimisation vs site/BESS                             & \cite{Driivz2026, Ampeco2026, Monta2026, OceanEV2026, Greenflux2026, LastMileSolution2026, Vitra2026, Current2026, Emabler2026, ChargePoint2026, EOCharging2026, ChargeLab2026, TridensTechnology2026, ChargePanel2026, Reev2026, Evoltsoft2026, Vialumina2026}                                          \\ \hline
Price-based optimisation (day-ahead)               & High                                            & Shift power within tariff windows                 & Tariff feeds (TOU/RTP)                                             & Conflicts with route SLAs                                & \cite{Driivz2026, Ampeco2026, Monta2026, OceanEV2026, Greenflux2026, LastMileSolution2026, Vitra2026, Emabler2026, ChargePoint2026, TridensTechnology2026, Reev2026, Vialumina2026}                                         \\ \hline
Ancillary service market participation             & Limited                                         &  Adoption of the OpenADR standard                                               &    OpenADR control signal                                                               &   Hardware constraints, clashes with operational fleet priorities                                                 & \cite{Driivz2026, Ampeco2026, Monta2026, LastMileSolution2026, Vitra2026, Emabler2026, ChargeLab2026}                                          \\ \hline
Reservation / Book-\&-Charge                       & High                                            & Slot booking via app/MSP                          & MSP/CPO platforms                                                  & No FMS linkage = planning blind                          & \cite{Driivz2026, Ampeco2026, Monta2026, Greenflux2026, LastMileSolution2026, Vitra2026, Emabler2026, ChargePoint2026, EOCharging2026, TridensTechnology2026, ChargePanel2026}                                          \\ \hline
Driver/user authentication \& billing              & High                                            & RFID/app; MSP clearing                            & Roaming hubs                                                       & Admin overhead for multi-tenant depots                   & \cite{Driivz2026, Ampeco2026, Monta2026, OceanEV2026, Greenflux2026, LastMileSolution2026, Vitra2026, Current2026, SwitchEV2026, Emabler2026, ChargePoint2026, EOCharging2026, ChargeLab2026, EVConnect2026, TridensTechnology2026, ChargePanel2026, Reev2026, Evoltsoft2026, ClenergyEV2026, Vialumina2026}                                           \\ \hline
Fleet schedule ingestion                           & Nascent                                         & Typically manual CSV/Excel upload or UI priorities          & FMS exports; operator inputs                                       & Stale data; error-prone                                  &  \cite{Driivz2026, Ampeco2026, ChargePoint2026, EOCharging2026, ChargeLab2026, TridensTechnology2026, ChargePanel2026}                                        \\ \hline
Solar forecast \& metreing integration             & Nascent                                        & DLM adjusted to PV surplus, no optimisation                        & PV metres/forecasts                                                & No formal EMS co-optimisation                            & \cite{Driivz2026, Ampeco2026, Monta2026, Greenflux2026, Reev2026}                                          \\ \hline
BESS coordination                                  & Nascent                                         & Indirect (Connect with EMS)                       & EMS signals (if any)                                               & Rare direct CPMS - BESS control                          & \cite{Driivz2026, Greenflux2026, Emabler2026}                                          \\ \hline
Charging demand forecasting                        & Nascent                                        & Statistical/ML on past arrivals                   & CPMS history; sometimes telematics                                 & Breaks under atypical ops                                & \cite{Greenflux2026, Emabler2026}                                          \\ \hline
Bi-directional charging optimisation               & Many Pilots                                    &  ISO-15118/20 communication                     &   Vehicle departure time, Tariff feeds                                         &    Limited hardware available, potential unknown battery round trip efficiencies (charging, discharging)              & \cite{Driivz2026, Ampeco2026, Monta2026, Greenflux2026, LastMileSolution2026, Vitra2026, Current2026, SwitchEV2026, Emabler2026, ChargePoint2026, TridensTechnology2026, ClenergyEV2026, Vialumina2026}                                                  \\ \hline
Supporting demand pooling (VPP)                    &   Emerging                                &   Bespoke API integrations between CPMS and VPP                       & VPP control signals                                                        &  Regulatory Complexity                                          &     \cite{Driivz2026, Ampeco2026, Monta2026, Greenflux2026, LastMileSolution2026, Vitra2026, Emabler2026}                                               \\ \hline
\end{tabular}%
}
\end{table}
\subsection{EMS}
\begin{table}[H]
\resizebox{\textwidth}{!}{%
\begin{tabular}{|p{3cm}|p{1cm}|p{2.5cm}|p{3cm}|p{3cm}|p{1.5cm}|}
\hline
\textbf{Capability   / Decision}            & \textbf{Market   Prevalence} & \textbf{How It Is Done   Today}    & \textbf{Supporting data   (up- and downstream)} & \textbf{Observed   Limitations / Risks}      & \textbf{Reference   Companies} \\ \hline
PV dispatch                                                        & High                                              & Day-ahead \& intra-day; 15–60 min updates              & Weather forecasts, historical PV, irradiance sensors                 & Forecast error under fast cloud transients                        &\cite{nuvve_platform, RiDERgy2026, Synop2026, BluwaveAI2026, ABB2026, Ampcontrol2026, Ampowr2026, Honeywell2026, iWell2026, Schneider2026, eaton_brightlayer_energy}                                                \\ \hline
BESS scheduling (charge/discharge)                                 & High                                              & 15-min planning; real-time controller tracks setpoints & SoC, BESS constraints, tariffs, site load forecast                   & Siloed from charging priorities; no charger-aware co-optimisation & \cite{nuvve_platform, RiDERgy2026, Synop2026, BluwaveAI2026, ABB2026, Ampcontrol2026, Ampowr2026, Honeywell2026, iWell2026, Schneider2026, eaton_brightlayer_energy}                                            \\ \hline
Grid connection limit enforcement   (contracted capacity mgmt)     & High                                              & EMS enforces site cap; CPMS throttles locally          & Site metre, sub-metreing, breaker states                             & Sequential control can cause conservative throttling              & \cite{nuvve_platform, RiDERgy2026, Synop2026, BluwaveAI2026, ABB2026, Ampcontrol2026, Ampowr2026, Honeywell2026, iWell2026, Schneider2026, eaton_brightlayer_energy}                                            \\ \hline
Tariff/price optimisation (TOU / RTP /   spot)                     & High                                              & Day-ahead scheduling; intra-day re-optimisation        & Tariff feeds, RTP/spot prices, DSO peak charges                      & Objective conflicts with logistics SLA without fleet context      & \cite{nuvve_platform, RiDERgy2026, Synop2026, BluwaveAI2026, ABB2026, Ampcontrol2026, Ampowr2026, Honeywell2026, iWell2026, Schneider2026, eaton_brightlayer_energy}                                            \\ \hline
Demand response (DR) / ancillary markets                           & Low / Moderate                                      & Manual enrollment; limited auto-bidding                & Market signals, aggregator APIs                                      & Multi-market optimisation limited; service conflict risk          & \cite{nuvve_platform, BluwaveAI2026, ABB2026, Ampcontrol2026, Honeywell2026, Schneider2026, eaton_brightlayer_energy}                                            \\ \hline
Non-fleet flexbile load scheduling   (HVAC/process)                & High                                              & Statistical/ML forecast; 15–60 min                     & Weather, occupancy, historical                                       & Often not exposed to CPMS; blocks joint optimisation              & \cite{ABB2026, Honeywell2026, Schneider2026}                                            \\ \hline
Direct charger control (OCPP/ModBus   ingestion)                   & Low (rare)                                        & Limited to few EMS+CPMS hybrids                        & OCPP 1.6-J/2.0.1 session data                                        & Most EMS lack OCPP parsing; CPMS holds these data                 & \cite{nuvve_platform, RiDERgy2026, Synop2026, BluwaveAI2026, ABB2026, Ampcontrol2026, Ampowr2026, Honeywell2026, Schneider2026, eaton_brightlayer_energy}                                            \\ \hline
CPMS - EMS integration (indirect charger   control)                & Emerging (API only)                               & One-way via APIs; no bi-directional control            & CSV/API payloads (bookings, priorities, caps)                        & No unified optimiser → partially observable optimisation          & \cite{nuvve_platform, RiDERgy2026, Synop2026, BluwaveAI2026, ABB2026, Ampcontrol2026, Ampowr2026, Schneider2026, eaton_brightlayer_energy}                                            \\ \hline
Multi-fleet site operations (tenanting,   cross-billing awareness) & No evidence                    & CPMS handles booking/roaming; EMS unaware of tenants   & RFID/app IDs, MSP contracts                                          & EMS cannot attribute energy/costs per fleet without CPMS          & None                                            \\ \hline
Supporting demand/supply pooling (VPP)                             &    Nascent                                 &       Via bespoke VPP and EMS API integration              &   VPP control signals                                                &     Regulatory complexity                                      &    \cite{nuvve_platform, Synop2026, BluwaveAI2026, iWell2026, eaton_brightlayer_energy}                                                  \\ \hline
Fleet management integration                             &    Nascent                                 &       Via bespoke API integration              &   Fleet schedules                                                &     Conflict resolving on scarce assets                                      &    \cite{nuvve_platform, RiDERgy2026, Synop2026, BluwaveAI2026, Ampcontrol2026}                                                  \\ \hline
Bi-directional charging integration                             &    Nascent                                 &       Via ISO15118/20              &   Commands                                                &     Charging efficiency monitoring                                      &    \cite{nuvve_platform, Synop2026, BluwaveAI2026, Ampcontrol2026}                                                  \\ \hline
\end{tabular}%
}
\end{table}

\subsection{FMS}
\begin{table}[H]
\resizebox{\textwidth}{!}{%
\begin{tabular}{|p{2.5cm}|p{1.5cm}|p{3.5cm}|p{2cm}|p{2cm}|p{2.5cm}|}
\hline
\textbf{Capability   / Decision}                       & \textbf{Market   Prevalence} & \textbf{How It Is Done   Today} & \textbf{Supporting data   (up- and downstream)} & \textbf{Observed   Limitations / Risks} & \textbf{Reference   Companies} \\ \hline
Planning with EV range restriction                                            & Moderate-High                                     &   Apply static range limits from OEM or telematics historics before route assignment                                    & OEM/telematics range, static rules                                   & Over-conservatism, under-utilisation                         & \cite{MichelinConnectedFleet2026, KiaPBVFMS2026, SmartTrak2026, Volteras2026, ZeroMission2026, Fryte2026, BetterFleet2026, OptiBus2026, Katsana2026, ZonarSystems2026, MakeMyDay2026, AutoWiz2026, FleetX2026, TrinetraWireless2026, Loginext2026, GoMotive2026, FarEye2026, NextBillionAI2026, Iveco2026, PTVLogistics2026, Ortec2026, Navixy2026, Webfleet2026, TargaTelematics2026, Scania2026, Samsara2026, Here2026, chargetrip, tno_optimal_charging_logistics_2024}                                            \\ \hline
Planning with route-specific consumption   models (grade, temp, payload, HVAC) & Moderate                                          &  Either ML-based predictive models trained on historical telematics data or first-order physics-based models (grade, temperature, payload, HVAC) to estimate energy use per route.                                                   & Map/elevation, weather, payload, historical                          & Misestimated SoC                                             & \cite{Volteras2026, ZeroMission2026, BetterFleet2026, OptiBus2026, MakeMyDay2026, GoMotive2026, NextBillionAI2026, Iveco2026, PTVLogistics2026, Ortec2026, Here2026, chargetrip, tno_optimal_charging_logistics_2024}                                            \\ \hline
Integrated charging stop planning (depot)                                     & Rare                                              &  Rare; mostly conceptual. No automated logic                                                   & Charger locations, power, availability/prices                        & Feasible plans ignore charging                               & \cite{Volteras2026, ZeroMission2026, OptiBus2026, Katsana2026, ZonarSystems2026, MakeMyDay2026, AutoWiz2026, FleetX2026, TrinetraWireless2026, FarEye2026, NextBillionAI2026, Iveco2026, PTVLogistics2026, Navixy2026, TargaTelematics2026, Samsara2026, chargetrip, tno_optimal_charging_logistics_2024}                                            \\ \hline
Integrated charging stop planning   (public)                                  & Rare                                              &   Rare; mostly conceptual. No automated logic                                                  & Charger locations, power, availability/prices                                                                     &  Feasible plans ignore charging                                                            &  \cite{Fryte2026, Katsana2026, MakeMyDay2026, AutoWiz2026, FleetX2026, TrinetraWireless2026, FarEye2026, NextBillionAI2026, PTVLogistics2026, TargaTelematics2026, Samsara2026, Here2026, chargetrip, tno_optimal_charging_logistics_2024}                                                   \\ \hline
Dynamic re-routing with SoC constraints                                       & Rare                                              &  Trigger re-route when SoC < threshold using live traffic and charger status.                                       & Live traffic, SoC, charger status                                    & Inflexibility to disturbances                                & \cite{OptiBus2026, Samsara2026, Here2026, chargetrip, tno_optimal_charging_logistics_2024}                                            \\ \hline
Planning driver/legal shift constraints   within charging windows             & Rare                                              &  Not seen publicly yet                                                   & Labor rules, depot hours, charger booking                            & Compliance vs energy cost conflicts unmanaged                & \cite{OptiBus2026, FarEye2026}                                            \\ \hline
Book-and-Charge integration (reservation   in plan)                           & Rare                                              &  Done on a structural/manual bases, for highly predictable routes (i.e. longhaul shuttle services)                                       & CPMS/MSP booking APIs                                                & Queueing risk, missed SLAs                                   & \cite{SmartTrak2026, Fryte2026, MakeMyDay2026, Here2026}                                            \\ \hline
Price/CO2-aware planning                                                      & Nascent                                           &  Not seen yet                                                   & RTP/CO2 intensity feeds                                              & Cost/footprint not optimised                                 & \cite{FleetX2026, TrinetraWireless2026, FarEye2026, NextBillionAI2026, Navixy2026, tno_optimal_charging_logistics_2024}                                            \\ \hline
Planning with battery health-aware   objective                                & Nascent                                           &  Apply soft penalties for deep cycles (low SoC); heuristic scoring.                                                  & BMS data, cycle models                                               & Long-term TCO penalties                                      & \cite{FleetX2026, Webfleet2026, tno_optimal_charging_logistics_2024}                                            \\ \hline
\end{tabular}%
}
\end{table}

\end{document}